\documentclass[pre,aps,twocolumn]{revtex4}

\usepackage{graphicx,subfigure,amsbsy,amsmath,float}
\usepackage[latin1]{inputenc}

\font\bbfnt=msbm10

\def\bbR{\mbox{\bbfnt R}}

\begin{document}

\title{Peeling Bifurcations of Toroidal Chaotic Attractors}

\author{Christophe Letellier$^1$, Robert Gilmore$^{1,2}$,
and Timothy Jones$^{2}$} 

\affiliation{$^1$ CORIA UMR 6614 --- Universit\'e de Rouen, 
Av. de l'Universit\'e, 
BP 12, F-76801 Saint-Etienne du Rouvray cedex, France}

\affiliation{$^2$ Physics Department, Drexel University, Philadelphia,  
Pennsylvania 19104, USA}

\date{{\it Physical Review E}, submitted on July 12, 2007 
--- Revised on \today}

\begin{abstract}
Chaotic attractors with toroidal topology (van der Pol attractor)
have counterparts with symmetry that exhibit unfamiliar phenomena.  We
investigate double covers of toroidal attractors, discuss changes in
their morphology under correlated peeling bifurcations, describe
their topological structures and the changes undergone as a symmetry
axis crosses the original attractor, and indicate how the symbol name
of a trajectory in the original lifts to one in the cover.  Covering 
orbits are described using a powerful synthesis of kneading theory
with refinements of the circle map.  These methods
are applied to a simple version of the van der Pol oscillator.
\end{abstract}

\pacs{PACS numbers: 05.45.+b}

\maketitle

\section{Introduction}
\label{sec:introduction}

It has been known for some time that discrete symmetry
groups can be used to relate chaotic attractors with
different global topological structures
\cite{Mir93, Let01, Let03, Gil07a}.  By different (or distinct) 
topological structures we mean there is no smooth deformation of
the phase space that can be used to transform one
attractor into the other in a smooth way.  If a chaotic attractor has
a discrete symmetry, points in the attractor that are mapped
into each other under the discrete symmetry can be identified 
with a single point in an ``image'' attractor.  The identifications
are through local diffeomorphisms.  The original symmetric attractor
and its image are not globally topologically equivalent.  This process can be
run in reverse.  A chaotic attractor without symmetry can be
``lifted'' to a covering attractor with a discrete symmetry
following algorithmic procedures \cite{Mir93, Let01, Let03, Gil07a, Let07}.  

\begin{figure}[ht]
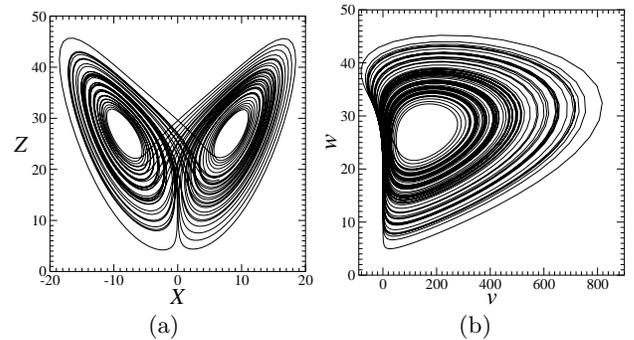

  \begin{center}
    \begin{tabular}{cc}
      \includegraphics[width=4.0cm]{covlor.eps} &
      \includegraphics[width=4.0cm]{imlor.eps}\\
       (a) & (b)  \\[-0.2cm]
    \end{tabular}
    \caption{The Lorenz attractor (a) can be mapped
to a R\"ossler-like attractor (b) by identifying points related
by rotation symmetry about the $Z$-axis.  This process is 
reversible:  R\"ossler-like attractors can be ``lifted'' to
Lorenz-like attractors.}
    \label{fig:mod_out}
  \end{center}
\end{figure}

A simple example illustrates these ideas.  The Lorenz attractor
\cite{Lor63} obtained with standard control parameter values
exhibits a two-fold symmetry.  The symmetry is generated by
rotations about the $Z$ axis through $\pi$ radians: $R_Z(\pi)$.  
We mod out this two-fold symmetry by identifying pairs of points 
$(X,Y,Z)$ and $(-X,-Y,Z)$ in the symmetric attractor
with a single point $(u,v,w)=(X^2-Y^2,2XY,Z)$ in the image attractor.
This results in a chaotic attractor that is not topologically
equivalent to the original attractor.  Rather, it is topologically
equivalent (not diffeomorphic, \cite{Mir93, Let01, Gil07a}) with the
R\"ossler attractor \cite{Ros76}.  Similarly, 
the R\"ossler attractor can be lifted to
a two-fold covering attractor that is topologically equivalent
to the Lorenz attractor.    The lift is carried out by inverting
the $2 \rightarrow 1$ local diffeomorphism used to generate the image:
$(u,v,w) \rightarrow (X = \pm \sqrt{ \frac{1}{2} (r+u)},
Y = \pm \sqrt{ \frac{1}{2} (r-u)},Z=w)$, where
$r = \sqrt{u^2+v^2}=X^2+Y^2$.  This modding out process is
illustrated in Fig. \ref{fig:mod_out}.

\begin{figure}[ht]
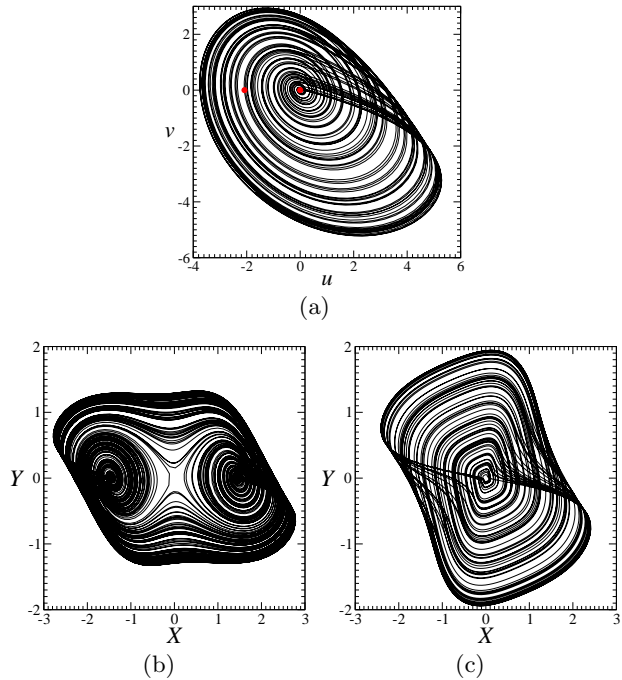

  \begin{center}
    \begin{tabular}{c}
       \includegraphics[width=4.0cm]{rosolo.eps} \\
       (a) \\[0.2cm]
    \end{tabular}
    \begin{tabular}{cc}
      \includegraphics[width=4.0cm]{rosfold3.eps} &
      \includegraphics[width=4.0cm]{rosfold1.eps} \\
      (b)   &  (c) \\[-0.2cm]
    \end{tabular}
    \caption{The R\"ossler attractor (a) can be lifted to
topologically distinct double covers with rotation symmetry
by placing the rotation axis in different positions:
(b) in the attractor; (c) in the hole in the middle
of the attractor.}
    \label{fig:lifts}
  \end{center}
\end{figure}

A single image attractor can have many topologically inequivalent covers,
all with the same symmetry group.  These covers are differentiated
by an index \cite{Let03,Gil07a,Let07}.  The index has interpretations at
the topological, algebraic, and group theoretical levels.  Briefly,
the index describes how the singular set of the local diffeomorphism
relating cover and image attractors is situated with respect to
the image attractor.  Different lifts of the R\"ossler attractor,
all with $R_Z(\pi)$ symmetry but with different indices, are obtained
if the two-fold rotation axis: passes through the hole in the
middle of the R\"ossler attractor; passes through the attractor itself; or 
passes outside both the attractor and the hole in the middle
(cf., Fig. \ref{fig:lifts}).
The transition of the symmetry axis through the attractor
is responsible for peeling bifurcations \cite{Let01}.

R\"ossler-like attractors have been lifted to covers with many
symmetry groups \cite{Let07}.  Whenever the singular set of the symmetry group
involves a rotation axis, this axis has passed through the attractor
at most once in all previous studies.  More precisely, it has 
passed through the branched manifold 
\cite{Gil98, Gil02,Bir83a, Bir83b} 
describing the attractor at a single point.

There have been no studies of covering attractors that are obtained
when the rotation axis intersects the image attractor in more
than one spot.  Generically, a rotation axis must intersect
a toroidal attractor an even number of times.  In the present work
we look at two-fold covers of toroidal attractors with rotation
symmetry.  We use methods similar to those used in \cite{Mir93, Let01,
Let03, Gil07a, Let07}.  Many of our results depend on a powerful synthesis
of kneading theory with refinements of the circle map.

In Sect. \ref{sec:vdP} below we study peeling bifurcations
for rotation-symmetric lifts of a chaotic attractor
generated by the van der Pol equations subject to a periodic drive.  
The phase space of this attractor is a ``hollow donut'' or ``fat tire'', 
that is, the direct product of an annulus with a circle: $A^2 \times S^1$.  
An annulus itself is a circle with a thick circumference:
$A^2 = S^1 \times I$, where $I$ is an interval, or short segment
of the real line.  The branched manifold that describes this
attractor is essentially a torus (thin tire).  A rotation
axis used to construct symmetric lifts must intersect the 
torus an even number of times.  Lifts of laminar and
chaotic flows on a torus are discussed in Secs. \ref{sec:rotations}
and \ref{sec:chaotic_attractors}, respectively.  The first of these
sections describes how the symbolic dynamics of image trajectories
lift to symbolic dynamics in covering trajectories.  The second
describes how the branched manifold describing the image attractor
lifts to the branched manifold describing the covering attractor.
We prepare for these discussions by introducing toroidal coordinates 
and describing flows on a torus in Sec. \ref{sec:flows_torus}.
We begin this entire odyssey in Sec. \ref{sec:peeling} with a review 
of peeling bifurcations and their properties.  Our results are
summarized in Sec. \ref{sec:summary}.

\section{Review of Peeling Bifurcations}
\label{sec:peeling}

Peeling bifurcations arise naturally when considering
covers of chaotic attractors.   They describe the bifurcations
these covers can undergo as the relative position of the
image attractor and the symmetry axis changes.  
Peeling bifurcations have been described in
some detail for covers of the R\"ossler dynamical system
in \cite{Let01, Gil07a}.  

We briefly describe the basic
idea for double covers with $R_Z(\pi)$ symmetry
about a rotation axis ${\cal R}$ with the usual saddle-type
symmetry.  Lift an image attractor (Fig. \ref{fig:lifts}(a))
to a double cover with 
${\cal R}$ far away from the original image attractor.
The double cover consists of two identical copies of the
original image attractor.  They are disconnected.  An
initial condition in one will evolve on that attractor
for all future times in the absence of noise.
As the rotation axis ${\cal R}$
approaches the image the two disjoint components of the
double cover approach each other, keeping ${\cal R}$
between them.  At some point ${\cal R}$ will intersect
the attractor.  When this occurs the rotation axis will
split the outer edge of the flow from one of the
two components of the cover and send it to the other
component, and {\it vice versa}.  The two attractors in the cover are no
longer disconnected (Fig. \ref{fig:lifts}(b)).  
As the rotation axis ${\cal R}$
moves deeper to the center of the image attractor (towards
the center of rotation of the R\"ossler attractor,
for example), the double cover becomes smaller in spatial
extent.  Finally, the rotation axis ${\cal R}$ may stop
intersecting the image by passing into the hole in the
middle (Fig. \ref{fig:lifts}(c)).  

The peeling bifurcation takes place as the
rotation axis moves from the outside to the inside of the
image attractor.  The image attractor itself is not
affected.  All bifurcations take place in the cover.
Before intersections begin and after
they end the double covers are structurally stable
and topologically inequivalent.
During the intersection phase the cover is 
structurally unstable because slight changes in the 
position of the rotation axis produce profound changes
in lifts of trajectories from the image. 
That is, a trajectory remains unchanged in the image
while its lift(s) into the cover change dramatically
as the rotation axis ${\cal R}$ moves. Changes in the shape, 
structure, and  period of lifts of unstable periodic 
orbits from the image  into the cover are predictable.  
These changes are summarized for covers of R\"ossler-like 
attractors in Figs. 7 - 12 of ref.  \cite{Let01}.

It may be useful to regard peeling bifurcations in terms
of how two trajectories in the cover that result from
the lift of a single trajectory in the image connect or
reconnect as the position of the symmetry axis changes.
Roughly but accurately, they can `turn back' into the
subset of the attractor from which they originated
when the axis is outside the trajectory, or else
`cross over' into the complement of that subset when the
axis is inside the trajectory.  This behavior is
reminiscent of what happens during the transition of a plane
through a saddle point on a surface, with the added
feature of `direction'.  Since orbits are dense in a
strange attractor, the lifted system cannot be structurally
stable during a peeling bifurcation.

\section{Flows on a Torus}
\label{sec:flows_torus}

\begin{figure}[ht]
  \begin{center}
    \includegraphics[width=8.0cm]{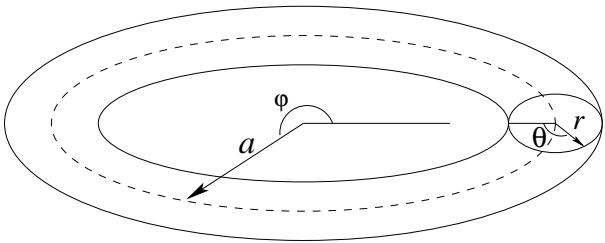} \\[-0.2cm]
    \caption{Toroidal coordinates.}
    \label{fig:torus}
  \end{center}
\end{figure}

It is useful to describe flows on a torus in terms of
a system of coordinates adapted to the torus: 
$(\phi, \theta, r)$.  In such a coordinate system 
$\phi$ is the longitude; it increases with time:
$d \phi/dt > 0$.  The angle $ \theta$ is the
meridional angle, measured from ``the inside of the torus''
(see Fig. \ref{fig:torus}), and $r$ measures the distance from
the center line of the torus, a circle of radius
$\rho$ in the $x$-$y$ plane.  The circle radius
$\rho$ must be sufficiently large so that $\rho-r > 0$ 
for all points in the attractor.  Standard cartesian coordinates
are represented in terms of these toroidal coordinates by
\begin{equation}
  \begin{array}{rcl}
    x &=& (\rho - r \cos \theta)\cos \phi \\
    y &=& (\rho - r \cos \theta)\sin \phi \\
    z &=& -r \sin \theta
  \end{array}
\label{eq:inject}
\end{equation}

The Birman-Williams theorem \cite{Gil98,Gil02,Bir83a, Bir83b} can be
applied to dissipative toroidal flows in $\bbR^3$ that generate
strange attractors.  The result is that the topology of the
flow is described  by a branched manifold.  The mechanism 
generating chaotic behavior involves an even number of folds.
The branch ``lines'' are now circles.  Since
the flow occurs in a bounding torus \cite{Tsa03, Tsa04}
of genus one,
the Poincar\'e surface of section consists of a single disk.
The intersection of the branched manifold with the disk
(``branch line'') is topologically a circle, $S^1$.
As a result, the flow can be investigated by studying maps of the
circle to itself \cite{Lan87}.

\section{Lifts of Rigid Rotations}
\label{sec:rotations}

In order to determine the topological structure of a strange
attractor in $\bbR^3$ it is sufficient to determine the topological
structure of the branched manifold that describes it.  This remains
true for covers of strange attractors with arbitrary symmetry 
\cite{Gil07a, Let07,Gil98}.  We do this in the following section.

\begin{figure}[ht]
  \begin{center}
    \includegraphics[width=7.5cm]{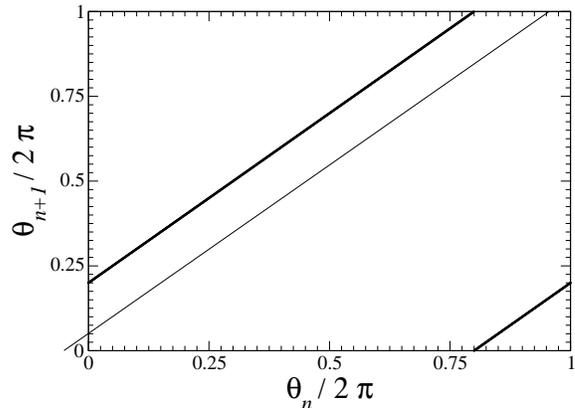} \\[-0.4cm]
    \caption{The return map for a rigid rotation is a straight line 
mod $2 \pi$. $\Omega/2\pi = 0.20$.}
    \label{fig:rigid_return}
  \end{center}
\end{figure}

In this section we prepare the groundwork by investigating how
a rigid rotational (quasiperiodic) flow on a torus is lifted to 
a double cover of the torus.  This is easily done by setting
$\rho=2, r=1, \theta = \alpha \phi$ in the toroidal coordinates. This
curve closes or does not close depending on whether $\alpha$
is rational or irrational.  The return map on a plane
$\phi =$ const. is shown in Fig. \ref{fig:rigid_return}.
It is $\theta_{n+1} = \theta_n +\Omega$ mod $2 \pi$.

\begin{figure}[ht]
  \begin{center}
    \includegraphics[width=7.0cm]{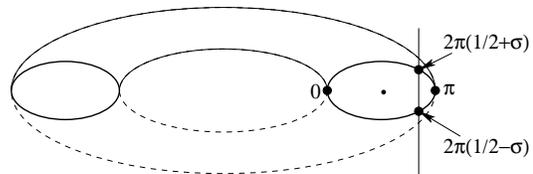} \\[-0.4cm]
    \caption{The order-two rotation axis intersects the 
torus at $2 \pi( \frac{1}{2} \pm \sigma) $.}
    \label{fig:torus_Z}
  \end{center}
\end{figure}

\begin{figure}[ht]
  \begin{center}
    \includegraphics[width=7.0cm]{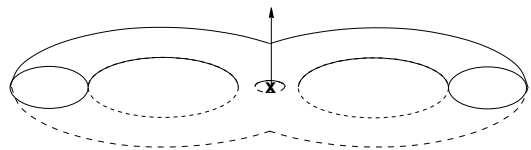} \\[-0.3cm]
    \caption{When the order-two rotation axis intersects the torus
at $2 \pi(\frac{1}{2} \pm \sigma)$, the double cover consists
of a structure with the topology shown.  This geometric structure
exists inside a bounding torus of genus three.}
    \label{fig:split_torus}
  \end{center}
\end{figure}

Now pass a rotation axis through the torus as shown in 
Fig. \ref{fig:torus_Z}.  The order-two rotation axis 
intersects the torus in an interval $I$.  If the 
rotation axis is parallel to the $Z$ axis, the end
points of this interval are at $\theta = 2 \pi (\frac{1}{2} \pm \sigma)$.
The rotation symmetry lifts the torus into a structure 
inside a genus-three bounding torus that is shown in 
Fig. \ref{fig:split_torus}.  The location of the rotation 
axis is indicated by $\times$.

\begin{figure}[ht]
  \begin{center}
    \includegraphics[width=7.0cm]{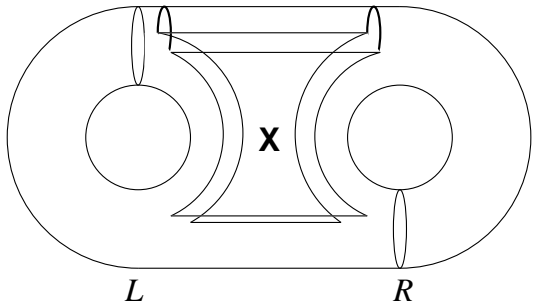} \\
    \caption{The laminar flow on a torus lifts
to a flow  on a manifold with the complicated form shown. 
The rotation axis is indicated by cross.}
    \label{fig:torus3}
  \end{center}
\end{figure}

\begin{figure}[ht]
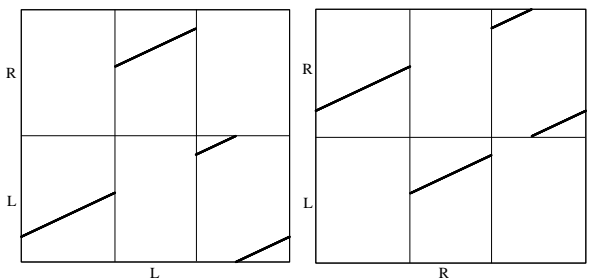

  \begin{center}
    \begin{tabular}{cc}
      \includegraphics[width=3.8cm]{rigid_L.eps} &
      \includegraphics[width=3.8cm]{rigid_R.eps} \\[-0.3cm]
    \end{tabular}
    \caption{Return map for a rigid flow contained within a genus-three
torus consists of mappings from two branch circles
to themselves.}
    \label{fig:return_2}
  \end{center}
\end{figure}

Since the flow exists in a bounding torus of genus three
the global Poincar\'e surface of section has two components
\cite{Tsa03, Tsa04}.
In such cases the first return map consists of a $2 \times 2$
array of maps \cite{Gil07a, Let05}.  For the Lorenz attractor
such maps indicate flows from branch line to branch line.
In the present case the return map indicates flows from the branch
circles on the left and right of Fig. \ref{fig:torus3}.
The return map for the cover of the rigid rotational flow,
in the case that the $Z$ axis intersects the image torus
at $\theta = \pi \pm 2 \pi \sigma$, is presented in 
Fig. \ref{fig:return_2}.  The angles parameterizing
the branch circles on the left and right run from
zero to $2 \pi$.  This figure shows that an
initial condition within $2 \pi \sigma$ of $\theta = \pi$
on the left hand branch circle maps to the right hand
branch circle (see Fig. \ref{fig:return_2}(a)), and {\it vice versa}.

A symbol name for any trajectory on the covering flow is
easily constructed.  Assume the return map in the image
is $\theta_{n+1} = \theta_n + \Omega$ mod $2 \pi$.  Then $\theta_n = \theta_0
+ n \Omega$ mod $2 \pi$.  Write out this string of real numbers
and replace each value $\theta_n$ by 1 if $\theta_n
\in I$, zero otherwise.  This results in a string of symbols, for
example 00000 11111 00000... for initial condition $L$.  
Choose an initial condition
$L$ or $R$, for one side of the cover or the other.
Then repeat this letter following symbol 0, conjugate
this letter ($L \rightarrow R, R \rightarrow L)$ following
symbol 1.  This algorithm leads to
$00000 11111 00000.. \rightarrow LLLLL RLRLR RRRRR..$.
A rotation-symmetric trajectory has a conjugate sequence.

Depending on parameter values (e.g., $\alpha \ll \sigma <
\frac{1}{4}$) the symbol sequence can consist of 
long strings of $L$s, long strings of $R$s, and long
strings of $LR$s, giving the appearance of prolonged
rotation about three centers: one being the left-hand
torus in the lift, another being the right-hand torus in the lift,
and the third alternation about both in sequence when
$\theta_n$ falls in the interval $I$ over a large 
range of successive interations.

As the rotation axis sweeps from the outside to the
inside of the torus, the value of $\sigma$ increases.
The circular intervals for which transition from one 
side to the other takes place increases,
with the return map becoming more and more off-diagonal.
As the $Z$ axis approaches the inner part of the image
torus, the measure of $\theta$ values that map to the
same branch circle decreases, and becomes zero
when tangency occurs ($\sigma = \pi$).  At this point
the return map is completely off-diagonal.  This is an 
indication that the global Poincar\'e surface of section
is no longer the union of two disjoint disks.  A single
disk suffices.  This signals that the flow returns
to a flow of genus-one type, and the return map on the
single disk is $\theta_{n+1} = \theta_n +2 \Omega$ mod $2 \pi$
(notice the factor of 2).

In the limit when the rotation axis is outside the
torus (``$\sigma < 0$'') the double cover consists of
two disconnected tori.  An initial condition in one torus
remains forever in that torus.
In the limit when the rotation axis is in the
hole in the middle of the torus (``$\sigma > \frac{1}{2}$'') 
the double cover consists of
a single torus.  When the rotation axis goes through the origin
of cartesian coordinates, the longitudinal angle $\phi$
in the image increases twice as fast as the longitudinal angle $\Phi$
in the cover.

Simulations of peeling bifurcations for double- and triple-covers
of laminar flows on a torus can be found at \cite{Jon07}.

\section{Lifts of Chaotic Attractors}
\label{sec:chaotic_attractors}

There is a class of strange attractors, such as the
van der Pol attractor that we discuss in the following section,
whose phase space is a hollow donut, topologically
$A^2 \times S^1$, where $S^1$ describes the longitudinal
(flow) direction.  The intersection of the attractor with
a constant phase plane $\phi = $cst. occurs in an annulus
$A^2 = S^1 \times I$, where this $S^1$ describes the
meridional direction and $I$ is a small interval.
Under the Birman-Williams projection \cite{Bir83a, Bir83b}
the intersection
of the projected attractor with the plane $\phi = $cst.
is topologically a circle $S^1$.  The forward time map
$\phi \rightarrow \phi + 2 \pi$ is therefore a mapping
of the circle to itself.  As a result, many of the
properties of this class of attractors are determined 
by the properties of the circle map.  For example, the
number and labeling of the branches of the attractor's
branched manifold are determined by the appropriate
circle map.

\begin{figure}[ht]
  \begin{center}
    \includegraphics[width=8.0cm]{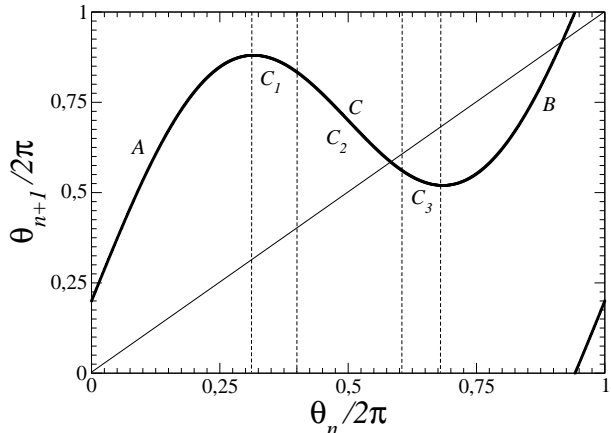} \\[-0.4cm]
    \caption{Circle map Eq. (\ref{eq:circle_map}) for $K=2.5, \Omega/2\pi = 0.2$.  
The three branches are conveniently labeled $A, C, B$.}
    \label{fig:circle_map}
  \end{center}
\end{figure}

For simplicity we assume that three branches 
$A, C, B$ suffice to describe the branched manifold, 
and that the return map
on the singularity at which these branches are joined
(the branch circle) is a circle map \cite{Lan87}:
\begin{equation}
  \theta_{n+1} = \theta_n + \Omega + K \sin \theta_n~~~{\rm mod~~}2\pi
  \label{eq:circle_map}
\end{equation}
with $K > 1$.  This map is shown in Fig. \ref{fig:circle_map}.
Branch $C$ is orientation reversing.  Its forward image
extends over a range less than $2 \pi$ and its extent is
delineated by the two critical points.
Branches $A$ and $B$ are orientation preserving and their forward
image extends over a range greater than $2 \pi$.  They are
delineated by the critical points that bound $C$ and the
inflection point between them.

\begin{figure}[ht]
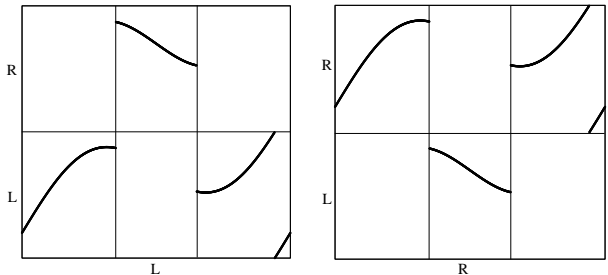

  \begin{center}
    \begin{tabular}{ccc}
      \includegraphics[width=3.8cm]{bicircle_L.eps} & ~ &
      \includegraphics[width=3.8cm]{bicircle_R.eps} \\[-0.2cm]
    \end{tabular}
    \caption{Return map for the double cover of the chaotic
flow whose return map is shown in Fig. \ref{fig:circle_map}.
Parameter value: $\sigma = 0.15$.}
    \label{fig:bicircle}
  \end{center}
\end{figure}

The return map for the double cover is obtained as in
the previous section.  We begin by looking at intersections
of the rotation axis near the outside of the torus, at
values $\theta = \pi \pm 2 \pi \sigma$, with
$\sigma$ small.  The return map
on the two branch circles is as shown in Fig. \ref{fig:bicircle}
for $\sigma = 0.15$.

The return map for the double cover is obtained from
the return map for the image as follows.  The vertical
lines through the maximum and the minimim and the
vertical axis at $\theta/2\pi = 0,1$ in Fig. \ref{fig:circle_map}
separate the return map into three branches $A,B,C$.

The two additional vertical lines separate branch $C$
into three branches: $C_3$ which is the interval $I$:
$\pi - 2 \pi \sigma < \theta < \pi + 2 \pi \sigma$,
and $C_1$ and $C_3$, which map $L \rightarrow L$ and
$R \rightarrow R$.  Generally there are no 
degeneracies, so these five vertical lines divide the circle
into five angular intervals.  In the present case both
endpoints of the interval $I$ occur inside the orientation-reversing
branch $C$, so this branch is divided into three
parts: $C_1, C_2, C_3$, as shown in Fig. \ref{fig:circle_map}.  
As a result, initial conditions
on branches $A$ and $B$, and the adjacent parts of branch
$C$, namely $C_1$ and $C_3$ of the circle on the left, map back
to that circle while initial conditions in the angular
interval $C_2$ on the left circle map to the right hand
circle.  The branched manifold describing the covering flow
has 10 branches with transitions summarized as follows:
\[  
  \begin{array}{cccc}
    L \rightarrow L & R \rightarrow R & L \rightarrow R & R 
\rightarrow L \\ \hline
    A & A &  &  \\
    C_1 & C_1 & &  \\
     & & C_2 & C_2 \\
    C_3 & C_3 & & \\
    B & B & & 
  \end{array}
\]
In the event that one (both) of the endpoints of the interval $I$
coincide with one (two) of the three points separating 
$A,C,B$ there are 8 (6) branches.  

Before the peeling bifurcation begins, when the two identical
covers are well separated, each is characterized by a branched
manifold with three branches.  At the end of the peeling
bifurcation, when the rotation axis ${\cal R}$ is inside the
image torus, there are  $9 = 3^2$ branches
that can be labeled $(A,C,B) \otimes (A,C,B)=
[AA,BA,CA,AB,BB,CB,CA,CB,CC]$ \cite{Let01, Let03}.  
Branches labeled by an
even number of letters $C$ are orientation preserving.

A periodic orbit in the image can be lifted to one or
two covering orbits.  The symbol name of the covering orbit
is obtained from the symbol name of the image orbit and
information about the interval $I$.  The name of the
orbit in the image is written out (e.g., $ABBCBBAC$ and 
refined according to the location of the interval $I$
(e.g., $ ABBC_2BBAC_3 $.  An initial condition
on the left or right ($L$ or $R$) is given and this is
changed whenever a trajectory passes through one of the
branches defined by $I$.  For example, under this algorithm
$ ABBCBBAC \rightarrow 
A_L B_L B_LC_{2L}B_LB_LA_L C_{3L}~~~
A_R B_R B_RC_{2R}B_RB_RA_R C_{3R}$.  A period-$p$ orbit
lifts to two period-$p$ orbits or one symmetric orbit of period
$2p$ depending on whether the image orbit maps through the
interval $I$ an even or odd number of times.  The
algorithm for lifting orbits from a toroidal flow
to a double cover involves a synthesis of kneading theory
with refinement of the circle map due to the intersection
of the rotation axis with the image toroidal flow.

\section{Application to the van der Pol Attractor}
\label{sec:vdP}

\begin{figure}[ht]
  \begin{center}
    \begin{tabular}{c}
      \includegraphics[width=7.0cm]{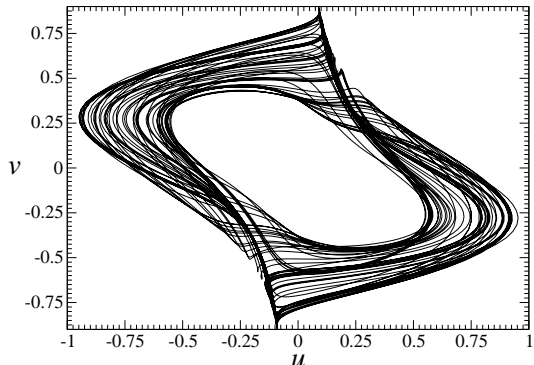} \\[-0.4cm]
    \end{tabular}
    \caption{Projections of the van der Pol attractor on the plane $u$-$v$.
Parameter values: $(a,b,c,d,\omega)=(0.25,0.7,1.0,10.0,\pi/2)$.  }
    \label{fig:vdp_attractor}
  \end{center}
\end{figure}

The Shaw version \cite{Sha81} version of the periodically driven van der Pol equations
\begin{equation} 
  \label{eq:vdp}
  \begin{array}{rcl}
    \dot{u} &=& bv+(c-dv^2)x \\
    \dot{v} &=&-u+a \sin(\omega t) 
  \end{array}
\end{equation}
produce a toroidal attractor for control parameter
values $(a,b,c,d,\omega)=(0.25,0.7,1.0,10.0,\pi/2)$
\cite{Gil07a,Gil02}.  The phase space for this attractor
is the direct product of an annular disk with a circle.
One projection of this attractor is shown in
Fig. \ref{fig:vdp_attractor}.

\begin{figure}[ht]
  \begin{center}
      \includegraphics[width=7.0cm]{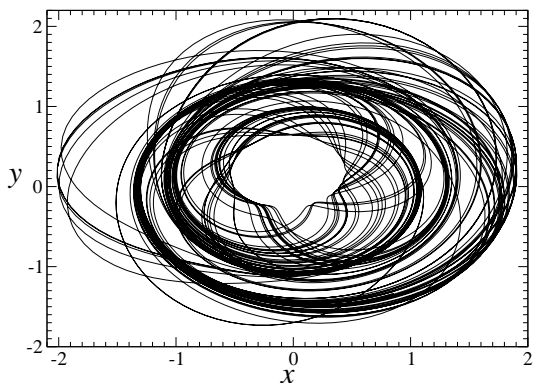} \\[-0.4cm]
    \caption{Mapping of the van der Pol attractor
into $R³$ using Eq. (\ref{eq:inject}) with $\rho = 1.2$.}
    \label{fig:vdp_image}
  \end{center}
\end{figure}

The attractor is mapped into $\bbR^3$ following the
prescription $x(t) = (\rho-u(t))\cos(\omega t),
y(t) =(\rho-u(t))\sin(\omega t), z(t) = v(t)$,
with $\rho=1.2$.  This flow, embedded in $\bbR^3$, has the 
topology of a hollow donut.  A projection onto the
$x$-$y$ plane is shown in Fig. \ref{fig:vdp_image}.

\begin{figure}[htbp]
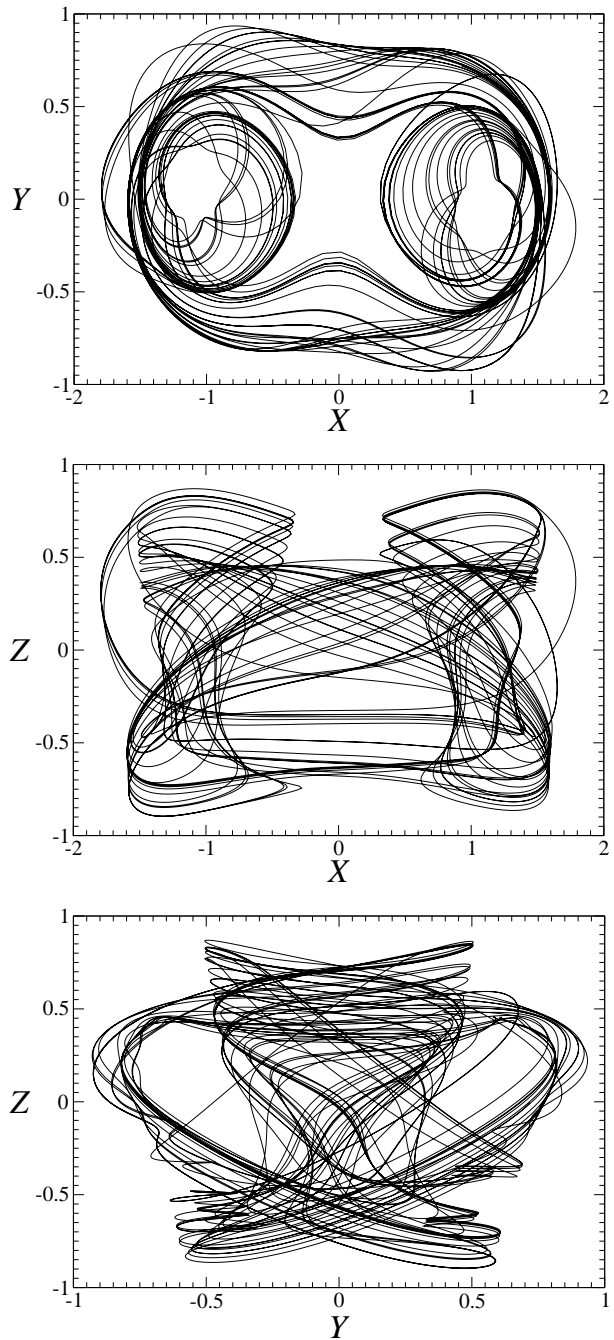

  \begin{center}
    \begin{tabular}{c}
      \includegraphics[width=8.0cm]{vdp2_xy.eps} \\[0.2cm]
      \includegraphics[width=8.0cm]{vdp2_xz.eps} \\[0.2cm]
      \includegraphics[width=8.0cm]{vdp2_yz.eps} \\[-0.2cm]
    \end{tabular}
    \caption{Double cover of the chaotic attractor 
solution of Equ.(\ref{eq:vdp}) 
for the Shaw version of the 
van der Pol equations is mapped from $D^2 \times S^1$ by a natural 
embedding.  The center line of the 
torus is mapped to a circle of radius 1.2 in the $x$-$y$ plane. 
A correlated peeling bifurcation 
occurs when the double cover is around the two-fold rotation axis 
through $(x,y)=(1.2,0.0)$.
The three images are projections onto the $X$-$Y$,
$X$-$Z$, and $Y$-$Z$ planes, from top to bottom.}
    \label{fig:vdp_peeling}
  \end{center}
\end{figure}

The covers of the chaotic attractor produced by 
this embedding into $\bbR^3$ undergo correlated peeling 
bifurcations as the rotation axis slices through
the image.  In Fig. \ref{fig:vdp_peeling} we show
three projections of the double cover obtained
when the symmetry axis is parallel to the $Z$
axis and has two-fold symmetry.  The rotation axis
passes through the point $(x,y)=(1.20,0.0)$ in the 
$x$-$y$ plane. The continuous version of this correlated
peeling bifurcation is available at \cite{Jon07}.

\begin{figure}[ht]
  \begin{center}
    \includegraphics[width=8.2cm]{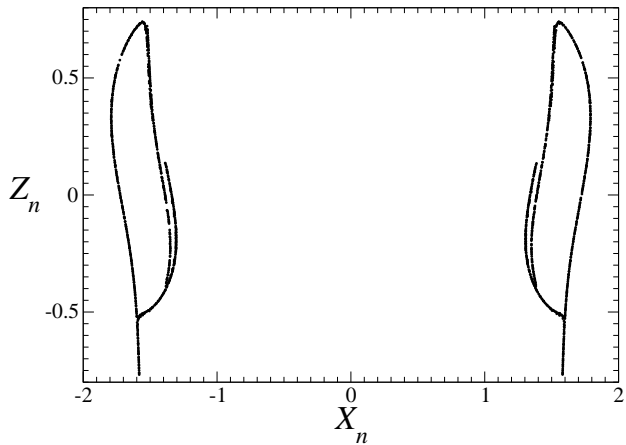}  \\[-0.3cm]
    \caption{Intersections of the double cover (Fig.\ \ref{fig:vdp_peeling}) 
of the van der Pol attractor with the two disconnected components
of the Poincar\'e section.
The component on the right has $Y=0,X>+1.1$ with $\dot{Y}>0$ and 
the component on the left has  $Y=0,X<-1.1$ with $\dot{Y}<0$.  
The rotation symmetry is clear.}
    \label{fig:ps_doublecover}
  \end{center}
\end{figure}

Both this attractor and the Lorenz attractor are contained
in genus-3 bounding tori.  The two attractors are
topologically inequivalent.
The $X$-$Y$ projection of this attractor, which is shown
in Fig. \ref{fig:vdp_peeling}(a), is similar to the
$X$-$Y$ projection of the Lorenz attractor.   However,
projections onto the other two directions are totally different.
The toroidal structure of the present attractor is revealed
in the projections shown in Figs. \ref{fig:vdp_peeling}(b)
and (c).  A Poincar\'e section of the double cover of the
van der Pol attractor (Fig. \ref{fig:ps_doublecover})
shows the double annular shape.  The two components
of the Poincar\'e surface of section consist of two
half planes, both with $Y=0$.  One is hinged on an 
axis parallel to the $Z$ axis through $(X,Y)=(1.1,0)$;
the other is the rotation image of the first.  Intersections
with $Y=0,\dot{Y}>0$ are taken on one half-plane and intersections
with $Y=0,\dot{Y}<0$ are taken with the other.  This Poincar\'e
section emphasizes the invariance of this attractor under
rotation symmetry around the $Z$ axis.

\section{Summary}
\label{sec:summary}

It is remarkable that the global topology of the image
attractor imposes nontrivial constraints on its
properties and those of its covers.  Specifically, an
attractor whose phase space is a hollow donut intersects a 
rotation axis an even number of times (more precisely, 
its branched manifold does).  Further, its branched 
manifold can have only an odd number of branches.
These remarkable properties extend, in a suitable way,
to double covers of these attractors.

For the first time methods for constructing double covers of chaotic
attractors have been applied to chaotic attractors
of a toroidal nature.  These attractors are
contained in genus-one bounding tori and are
described by branched manifolds with a circular cross 
section on a Poincar\'e surface of section \cite{Gil02}.  
Their return maps are maps of the
circle to itself.  Their double covers are created
by correlated peeling bifurcations.  The morphology of the
covering attractor changes systematically as the
rotation symmetry axis slices through the image torus
from outside to inside.  Outside, the double cover consists
of two identical attractors, each contained in a genus-one
torus. The two genus-one tori are disconnected.  
When the rotation axis intersects the image,
the double cover is contained in a genus-three torus and
is not structurally stable against
perturbations of the position of the rotation axis.
When the rotation axis enters the hole in the torus,
the double cover exists in a genus-one torus.
For various ranges of lift parameter values rotations can
appear to occur around a single center, two centers, or
three centers.  Lifts of periodic orbits in the
image attractor are described by a powerful synthesis of kneading theory
with refinements of the circle map.

Acknowledgement:  R. G. thanks the CNRS for an 
invited position at CORIA for 2006-2007.

\end{document}